
\magnification=1200
\def\sla{\raise.15ex\hbox{$/$}\kern-.57em}
\rightline{RU--92--20}
\vskip 2cm
\centerline {\bf Chiral Yukawa models in the planar limit.}
\vskip 1cm
\centerline {by}
\vskip .5cm
\centerline{George Bathas and Herbert Neuberger}
\vskip .5cm
\centerline{\it Department of Physics and Astronomy}
\centerline{\it Rutgers University}
\centerline{\it Piscataway, NJ 08855--0849}
\vskip 3cm
\centerline{\bf Abstract}
\vskip .5cm
We consider the most general renormalizable chiral Yukawa model with
$SU(3)_{\rm color}$ replaced by $SU(N_c)$, $SU(2)_{\rm L}$ replaced
by $SU(N_w )$ and $U(1)_{Y}$ replaced by
$U(1)^{N_w -1}$ in the limit
$N_c \rightarrow\infty$, $N_w \rightarrow\infty$
with the ratio $\rho=\sqrt{{N_w}\over{N_c}} \ne 0,\infty$ held fixed.
Since for $N_w \ge 3$ only one renormalizable Yukawa coupling per
family exists and there is no mixing between families
the limit is appropriate for
the description of the effects of a heavy top quark when
all the other fermions are taken to be massless. A rough
estimate of the triviality bound on the Yukawa
coupling is equivalent to $m_t \le 1~TeV$.
\vfill
\eject
In the minimal standard model at energies above 100$~ GeV $ only the
scalar selfcoupling $\lambda$
is completely unknown and might be strong. Skeptics\footnote{${}^1$}{The
assumption that the top Yukawa coupling is relatively
weak is consistent with present knowledge indicating that $m_t \le \sim 200
{}~GeV$.} might argue that a strong Yukawa coupling $y$
has not been yet ruled
out completely
and this note addresses the problem how to do calculations if
they are right [1].

Excepting the $\lambda$ and $y$ all other couplings are weak and can be
treated perturbatively. The problem then is to set up a computation first
in the case where all the weak couplings are set to zero. We have some
requirements: Because of ``triviality'' and ``vacuum stability'' issues
one ought to be able to keep an explicit cutoff in the calculations. The
cutoff has to be defined in a reasonable manner, meaning that an expansion
in inverse powers of the cutoff can be set up and has the normal structure
one would expect of terms induced by a higher embedding theory
(for example, a sharp momentum cutoff is not acceptable). We do
not want to use the lattice since many of the non--perturbative
effects in lattice models have little to do with continuum
physics.\footnote{${}^2$}{The lattice introduces in a smooth manner
a finite bottom to the Dirac
sea and this is particularly disturbing to chiral fermions. One could
however use our technique to study non--perturbative effects in lattice
models and hopefully sort out the situation there.} Intuitively
we feel that a sensible approximation should keep more or less a democratic
view of fermionic and scalar excitations.

We propose that a good way to achieve the above
is to consider a model with
a single Higgs multiplet where
$SU(3)_{\rm color}$ is replaced by $SU(N_c)$, $SU(2)_{\rm L}$ is replaced
by $SU(N_w )$ and $U(1)_{Y}$
is replaced by $U(1)^{N_w -1}$ in the limit
$N_c \rightarrow\infty$, $N_w \rightarrow\infty$
with the ratio $\bar\rho=\sqrt{{N_w}\over{2N_c}} \ne 0,\infty$ held fixed.
Just letting $N_c$($N_w$) go to infinity with $N_w$($N_c$) fixed suppresses
bosons(righthanded fermions) and these limits may distort effects like
vacuum destabilization [2] or bag formation [3]. In any case more diagrams
get summed up when both $N_c$ and $N_w$ go to infinity and all diagrams
you would get when only $N_c$ or $N_w$ go to infinity are included.

We introduce the model in Euclidean space; with a finite cutoff the
model is non-perturbatively well defined but the continuation to Minkowski
space would introduce unitarity violating effects at energies of the
order of the cutoff. This drawback is compensated by preserving translational
and full rotational invariance with propagators regulated in a Pauli--Villars
manner. The action is given by $S$:
$$
\eqalign{
-S=&\int_x\cr \bar \psi_{-}^{ia(r)}&\sla\partial_{+}\psi_{+}^{ia(r)}+
\bar\chi_{+}^{a(r)s} h_{\chi}(-\partial^2 )\sla\partial_{-} \chi_{-}^{a(r)s}+
\phi^{i*} h_{\phi}(-\partial^2 ) \partial^2 \phi^i -m_0^2 \phi^{i*}\phi^i -
{{\lambda_0}\over{4N_w}} (\phi^{i*}\phi^{i})^2\cr
+&{1\over{\sqrt{N}}}\sum_s [
\bar\psi_{-}^{ia(r)}\chi_{-}^{a(r^{\prime})s}\phi^{i}A_{rr^{\prime}}^{(s)}+
\bar\chi_{+}^{a(r)s}\psi_{+}^{ia(r^{\prime})}\phi^{i}A_{rr^{\prime}}^{*(s)}]
\cr}\eqno{(1)}
$$
where
$$
\eqalign{
N=&\sqrt{N_w N_c }, ~~~\sla\partial_{+}=\sigma_{\mu}\partial_{\mu},~~~
\sla\partial_{-}=\bar\sigma_{\mu}\partial_{\mu}\cr
\bar\sigma_1 =&\sigma_1 ,~~\bar\sigma_2 =\sigma_2 ,~~\bar\sigma_3 =\sigma_3 ,~~
\bar\sigma_4 =-\sigma_4 =-i\cr
i=&1,\cdots ,N_w~~~a=1,\cdots ,N_c~~~r=1,\cdots ,n_f~~~s=1,\cdots .N_w\cr
}\eqno{(2)}
$$
The $\sigma_{1,2,3}$ are Pauli matrices.
The functions $h_{\chi}$ and $h_{\phi}$
can be chosen in many ways and introduce the
regularization. Ultraviolet divergences can be cured by just regularizing the
$\chi$ and $\phi$ propagators because there are no loops consisting of
only $\psi$ propagators. When in need of an explicit form we shall use
$h_{\chi}(p^2 )=h_{\phi}(p^2 )=1+({p^2}/{\Lambda^2})^n$
with a sufficiently large $n$.

The symmetry properties are as follows: The space--time group is $SU(2)\times
SU(2)$ and a spinor with subscript $+$ transforms as $({1\over 2},0)$ while one
with subscript $-$ transforms as $(0,{1\over 2})$. $\bar\psi , \psi ,
\bar\chi , \chi$ are two component Grassmann spinors and complex
conjugation does not act on them. Under $SU(N_c )\times
SU(N_w )$ the representations are
 $\bar\psi_{-}\sim (\bar N_c , \bar N_w ),~\psi_{+} \sim
(N_c ,N_w ),~\bar\chi_{+} \sim (\bar N_c ,1),~\chi_{-}\sim(N_c ,1),~\phi^* \sim
(1,\bar N_w ),~\phi\sim (1,N_w )$. This representation content
is independent of the indices $r$ and $s$ whenever they appear;
the index $r$ runs over families and $s$
over the number of fermionic weak singlets
(and color mutiplets) per family.
The representations
under the remaining $U(1)$'s are chosen subject to anomaly considerations
when gauging is contemplated and eliminate all but one of the matrices
$A^{(s)}$. The non--vanishing matrix is chosen to be $A^{(1)}$ and the
remaining $\chi$ fields with index $s\ge 2$
decouple at zero gauge couplings. When $N_w =2$
an additional set of Yukawa couplings is allowed by $SU(N_c)\times SU(N_w )$
and also by the $U(1)$'s. These couplings do not generalize to
$N_w\ge 3$ and will henceforth be ignored.
The matrix $A_{rr^{\prime}}^{(1)}$ can be made diagonal with positive
entries by a bi--unitary transformation, decoupling
the families. In conclusion one can ignore the $r$ and $s$
indices, which is just as well because we
want to keep only one potentially heavy ``top'' color multiplet.
The model then becomes [1]:
$$
\eqalign{
-S=&\int_x  \bar \psi_{-}^{ia}\sla\partial_{+}\psi_{+}^{ia}+
\bar\chi_{+}^{a}h_{\chi}(-\partial^2 )\sla\partial_{-}\chi_{-}^{a}+
\phi^{i*}h_{\phi}(-\partial^2 )\partial^2 \phi^i -m_0^2 \phi^{i*}\phi^i -
{{\lambda_0}\over{4N_w}} (\phi^{i*}\phi^{i})^2\cr +&{g_0 \over{\sqrt{N}}}
[
\bar\psi_{-}^{ia}\chi_{-}^{a}\phi^{i}+
\bar\chi_{+}^{a}\psi_{+}^{ia}\phi^{i}];~~~~~~g_0 > 0.\cr
}\eqno{(3)}
$$

When $N\rightarrow\infty$ with $\bar\rho\equiv\sqrt{{N_w}\over{2N_c}}$
held fixed the dominating diagrams are planar. Renormalizability
allows us to get away with not having to introduce four Fermion
interactions and with them an insoluble planar diagram problem,
at no cost to generality.\footnote{${}^3$}{Except for possible
cutoff effects.} Thus the model is manageable being similar
diagrammatically
to two dimensional QCD with matter at large $N_c$. The large $N$ limit
can be found by looking at the structure of
planar Feynman graphs: essentially one sums over all
``cactuses of bubbles'' and all ``rainbows''. For future work it
is somewhat preferable to use functional integral manipulations to
the same end.

In this note we wish to show feasibility within a simplest nontrivial example.
We set our modest goal to compute a $\beta$--function associated
with the Yukawa coupling constant defined in some simple manner and ignoring
cutoff effects. This can be done with relative ease because even
at infinite order in the couplings the flow of this
Yukawa coupling is unaffected
by $\lambda_0$, similarly to the finite $N$ form at one loop order. From
this calculation a rough estimate for the ``triviality'' bound on the
top mass can be obtained.\footnote{${}^4$}{In practice this bound might be
somewhat uninteresting because it is likely to be higher than the
vacuum stability bound for reasonable Higgs masses.}

To find $\beta$ functions it suffices to solve the model at criticality
when all masses vanish. We start in the symmetric phase and pick a mass
independent renormalization scheme to make the zero mass limit smooth.

We introduce bilocal fields,
$$
S_{\alpha ,\beta}(x,y)
={1\over{N_c}} \bar\chi_{+\alpha}^a (x)\chi_{-\beta}^a (y),~~~~
K(x,y)={1\over{N_w}} \phi^{i*}(x)\phi^{i}(y),\eqno{(4)}$$
where $\alpha ,\beta$ are space--time spinor indices, into the functional
integral by writing representations for the appropriate $\delta$--functions
with the help of two auxiliary bilocal fields
$\lambda_{\alpha ,\beta} (x,y)$ and $\mu (x,y)$.
We can integrate out now all the original
fields, including the zero mode of the scalar field because we assumed that
we are in the symmetric phase. This makes the $N$ and $\bar\rho$ dependence
explicit and, when $N\rightarrow\infty$, the integral over
$S,\lambda ,K$ and $\mu$
is dominated by a saddle point which has to satisfy the following equations:
$$
\eqalign{
K(x,y)=&\{ [-h_{\phi}(-\partial^2 )\partial^2 + m_0^2 +\mu ]^{-1}\}_{y,x}\cr
S_{\alpha ,\beta}(x,y)=&\{ [h_{\chi}(-\partial^2 )\sla\partial_{-} +
\lambda ]^{-1}\}_{y\beta ,x\alpha}\cr
\lambda_{\alpha ,\beta}(x,y)=&-g^2 \bar\rho \sqrt{2}
\{\sla\partial_{+}^{-1}\}_{x\alpha ,y\beta } K(x,y)\cr
\mu (x,y) =&{{g^2}\over{\bar\rho\sqrt{2}}}\sum_{\alpha ,\beta}
\{\sla\partial_{+}^{-1}\}_{x\alpha ,y\beta } S_{\alpha ,\beta} (x,y)
+{{\lambda_0}
\over {2}} K(x,x)\delta_{x,y}\cr
}
\eqno{(5)}$$
We go to Fourier space and use translational and rotational invariance to write
$$
\tilde\lambda_{\alpha ,\beta} (p) =ip_{\mu}\bar\sigma_{\mu} \hat\lambda (p^2)~
{}~~~~~~~~\tilde\mu (p) =\hat\mu ( p^2 )
\eqno{(6)}$$
Let us introduce $Z_{\chi}$ and $Z_{\phi}$, wave function
renormalization constants, and new functions $f=Z_{\chi} [\hat\lambda (p^2 )
+h_{\chi} (p^2 )]$ and $g(p^2 )=Z_{\phi } [p^2 h_{\phi} (p^2 ) + m_0^2 +
\hat\mu (p^2 )]$. Since at the planar level there are no vertex corrections
to the Yukawa coupling\footnote{${}^5$}{This is also true at
one loop order at finite $N$.} it makes sense to define a renormalized Yukawa
coupling by $g_R^2 =Z_{\phi} Z_{\chi} g_0^2$, ($\alpha\equiv{{g_R^2}\over
{16\pi^2}}$). We also set $\lambda_0^{\prime} =\lambda_0 Z_{\phi}^2$ but,
unlike $\alpha$, do
not expect
$\lambda_0^{\prime}$ to stay finite when $\Lambda\rightarrow\infty$.
$Z_{\chi}$ and $Z_{\phi}$ are defined in terms of the bare couplings by
$f(\mu^2 )={{g(\mu^2 )}\over{\mu^2}}=1$.

After performing the angular part of the momentum integrals one gets
when the products in the saddle point equations for $x,y$ dependent quantities
are converted into convolutions in momentum space one finds:
$$
\eqalign{
f(p^2 )=&Z_{\chi}h_{\chi}(p^2 )+\sqrt{2}\alpha\bar\rho
\left [ {1\over 2}\int_{p^2}^{\infty}
{{dk^2 }\over{g(k^2 )}}
+{1\over{p^2 }}\int_0^{p^2 }
{{k^2 dk^2 }\over{g(k^2 )}}-
{1\over{2p^4 }}\int_0^{p^2 }
{{k^4 dk^2 }\over{g(k^2 )}} \right ]\cr
g(p^2 )=&Z_{\phi}p^2 h_{\phi}(p^2 )+{{\sqrt{2}\alpha}\over{\bar\rho}}
\left [ {{p^2}\over { 2}}\int_{p^2}^{\infty}
{{dk^2}\over{k^2 f(k^2 )}} +
\int_0^{p^2} {{dk^2}\over{f(k^2 )}}-
{1\over{2p^2 }}\int_0^{p^2}{{k^2 dk^2}\over{f(k^2 )}} \right ] +m^2 \cr
m^2=&Z_{\phi} m_0^2+{ {\lambda_0^{\prime}}\over{32\pi^2 }}
\int_0^{\infty} {{k^2 dk^2 }\over {g(k^2 ) }}-
{{\sqrt{2}\alpha}\over{\bar\rho}}
\int_0^{\infty}{{dk^2 }\over {f(k^2 )}}
\cr
}
\eqno{(7)}$$
The massless case is obtained by adjusting $m_0^2$ so that $m^2 =0$ and then
it makes sense to define $h(p^2 )={{g(p^2 )}\over {p^2}}$ leading to
a set of equations of a more symmetric appearance:
$$
\eqalign{
h(p^2 )=& Z_{\phi} h_{\phi} (p^2 ) +{{\sqrt{2}\alpha}\over{\bar\rho}}{\cal L}
[f](p^ 2)\cr
f(p^2 )=& Z_{\chi} h_{\chi} (p^2 ) +\sqrt{2}\alpha\bar\rho {\cal L}[h](p^2 )\cr
{\cal L}[X](p^2 )=&{1\over 2}\int_{p^2}^{\infty} {{dk^2}\over{k^2 X(k^2 )}}
+{1\over{p^2}}\int_0^{p^2}{{dk^2}\over{X(k^2 )}}-{1\over{2p^4 }}\int_0^{p^2 }
{{k^2 dk^2}\over{X(k^2 )}}\cr
f(\mu^2 )=&h(\mu^2 ) =1\cr
}
\eqno{(8)}$$
Note the interesting bosonic--fermionic symmetry at $\bar\rho =1$.
We now rescale the functions and the momentum variable by:
$$
f(u\mu^2 )=\sqrt{2\alpha } \bar\phi( u)~~~~~~~~h(u\mu^2 )=
\sqrt{\alpha } \bar\psi( u)
\eqno{(9)}$$
and obtain:
$$
\eqalign{
\bar\psi (u) =&{1\over{\sqrt{\alpha}}}+{{Z_{\phi}}\over{\sqrt{\alpha}}}
[h_{\phi} (u\mu^2 )-h_{\phi} (\mu^2 )]+ {1\over{\bar\rho}}\int_u^1 {{dv}\over
{v}}
\int_0^1
{{(1-x)dx}\over{\bar\phi (xv)}}\cr
\bar\phi (u) =&{1\over{2\sqrt{\alpha}}}+{{Z_{\chi}}\over{2\sqrt{\alpha}}}
[h_{\chi} (u\mu^2 )-h_{\chi} (\mu^2 )]+ \bar\rho\int_u^1 {{dv}\over {v}}
\int_0^1
{{(1-x)dx}\over{\bar\psi (xv)}}\cr
}
\eqno{(10)}$$

It is clear that when $n$ is large enough $h$ and $f$ are dominated by
the $h$ functions in the ultraviolet and therefore the large cutoff
behavior of the wave function renormalization constants is such
that in the infinite cutoff limit at $u$ and $\mu^2$ fixed
$Z_{\phi}$ and $Z_{\chi}$ simply disappear
from the above equation leaving us with
$$
\eqalign{
\bar\psi (u) =&{1\over{\sqrt{\alpha}}}+
{1\over{\bar\rho}} \int_u^1 {{dv}\over {v}} \int_0^1
{{(1-x)dx}\over{\bar\phi (xv)}} \cr
\bar\phi (u) =&{1\over{2\sqrt{\alpha}}}+
\bar\rho\int_u^1 {{dv}\over {v}} \int_0^1
{{(1-x)dx}\over{\bar\psi (xv)}}\cr
}
\eqno{(11)}$$

On physical grounds it is obvious that we wish that $\bar\psi$ and $\bar\phi$
be positive. The equations above show that if this is true in some interval
$(0,u_{*})$ then the functions will be monotonically decreasing there.
However, the equation
do not admit an asymptotic behavior as $u\rightarrow\infty$ with
both functions approaching non--negative limits so the positivity requirement
must get violated somewhere in the
ultraviolet.\footnote{${}^6$}{A zero of $\bar\psi$ or $\bar\phi$ at a positive
$u$ corresponds to a pole in a two--point function
at an Euclidean momentum; the ``particle'' associated
with this pole would be tachyonic.} Starting at the scale where the
violation first occurs cutoff effects cannot be neglected any more. This
``unphysical'' scale is the usual Landau pole, this time appearing in a
nonperturbative approximation.

In the infrared there is no Landau pole problem and
cutoff effects are indeed negligible. To investigate the behavior there
we set $x=-\log (u)$ and $ \hat\psi (x) =\bar\psi (u),~
\hat\phi (x) =\bar\phi (u) $ and derive:
$$
\eqalign{
{\cal D}\hat\psi =&{1\over{\bar\rho\hat\phi}}\cr
{\cal D}\hat\phi =&{{\bar\rho}\over{\hat\psi}}\cr
{\cal D}=&{{d^3}\over{dx^3}}-3{{d^2}\over{dx^2}}+2{{d}\over{dx}}={{d}\over{dx}}
\left ( {{d}\over{dx}} -1\right ) \left ({{d}\over{dx}} -2 \right ) \cr
}
\eqno{(12)}$$
These equations allow us to extract the infrared behavior of the solution
of the integral equations. We ended up with only differential
equations (rather than integral)
reflecting that according to the Renormalization Group
one needs only very limited information at a given scale in order
to derive the behavior at a scale close by; therefore a differential
equation must show up eventually, its order less the number
of asymptotic conditions determining the
amount of information at a given scale that is needed to determine
the behavior at the next scale. Our choice of regularization was
made so that even at finite cutoff $\hat\phi$ and $\hat\psi$ obey
a set of purely differential equations of a structure similar to (12).

Since the fixed point governing
the infrared behavior is the free field fixed point the form of the
solutions in the infrared simply embodies the two anomalous dimensions
associated with the $\chi$ and $\phi$ fields when
expanded in $\alpha$ around $\alpha =0$. These anomalous dimensions
as a function of the coupling also determine the $\beta$ function and
from the asymptotic series of $\hat\psi (x)$
and $\hat\phi (x)$ at $x\rightarrow\infty$ we can get the contributions
to the $\beta (\alpha )$ function ordered in the number of loops.

The differential equations lead to:
$$
\eqalign{
\hat\psi\sim &\psi_0 x^b [1+c{{\log x}\over x}+c^{\prime} {1\over x}+\cdots]\cr
\hat\phi\sim &\phi_0 x^a [1+d{{\log x}\over x}+d^{\prime} {1\over x}+\cdots]\cr
}
\eqno{(13)}$$
where
$$
\eqalign{
a=&b\bar\rho^2 ={{\bar\rho^2}\over{1+\bar\rho^2}}\cr
d=&c\bar\rho^2 =-{{3\bar\rho^4}\over{(1+\bar\rho )^3}}\cr
{{d^{\prime}}\over{\bar\rho^2 }}-c^{\prime} =&{{3(1-\bar\rho^2 )}\over{
2(1+\bar\rho^2 )^2 }};~~~~~\psi_0\phi_0 ={{\bar\rho +\bar\rho^{-1}}\over 2}\cr
}
\eqno{(14)}$$
Only the product $\psi_0\phi_0$ and a particular linear combination of
$d^{\prime}$ and $c^{\prime}$ get determined because
the equations are invariant
under a field rescaling $\hat\psi\rightarrow{A^{-1}}\hat\psi$,
$\hat\phi\rightarrow A \hat\phi$ reflecting a change in the finite parts of
the wave function renormalization constants and a
shift $x\rightarrow x+x_0$ reflecting a change in $\mu^2$. The
field--rescaling
invariance disappears in the product
$r(u)\equiv\sqrt{2} \bar\psi (u)\bar\phi (u)$
and this combination is indeed special because it satisfies a
Renormalization Group
equation in which only the $\beta$--function appears (the sum
of the two anomalous dimensions rather than each individual one):
$$
\left [ -{{\partial}\over{\partial t}}+\beta (\alpha )
{{\partial}\over{\partial\alpha}} \right ] r(e^{2t} ,\alpha )=0
\eqno{(15)}$$

This equation is exact in our limit. From it we derive another exact relation
between $r$ and $\beta$:
$$
\beta (\alpha (u\mu^2) )=
-2[\alpha (u\mu^2 )]^2
{{dr(u)}\over{d(\log u)}}{\bigg |}_{r(u)={1\over{\alpha (u\mu^2 )}}}
\eqno{(16)}$$
The argument of $\alpha$ in the equation above can of course be ignored.

Using now the asymptotic expansion of the exact
solution in the infrared we get:
$$
\beta(\alpha )=\alpha^2 [\sqrt{2} (\bar\rho +\bar\rho^{-1})-3\alpha+O(\alpha^2
)]
\eqno{(17)}$$
The one loop result when expressed in a conventional variable, $y(t)$
(in the standard model the top mass at tree level is $m_t =y(\sim 0)~
246~[GeV]$),
is, with the true value $\bar\rho={1\over{\sqrt{3}}}$, ${{dy^2}\over
{dt}}={{y^4}\over{\pi^2}}$ which is close to the finite $N$ result,
${{dy^2}\over{dt}}={{9y^4}\over{8\pi^2}}$. The difference comes from the
wave function renormalization of the lefthanded $\psi$ field which is
suppressed at leading order in ${1\over N}$. For arbitrary $N_c$ and $N_w$
we would have gotten at one loop order
${{dy^2}\over{dt}}={{(2N_c +N_w +1) y^4}\over{8\pi^2}}$
and we see that the relative magnitude of the
$1/N$ correction is much smaller than in the scalar case [4].

To get the full $\beta$--function at infinite $N$ we solve for $\hat\psi$
and $\hat\phi$ numerically by iterating an integral form of the
equations based on (7)
but with the normalization conditions at $\mu^2$
incorporated; this avoids the appearance of
double integrals like in (11-12). The procedure is probably safer than
using the third order differential equations and trying to enforce boundary
conditions at $x\rightarrow\infty$ that eliminate the exponentially growing
components $e^{x}$ and $e^{2x}$ (see the last line in equation (12)).
The result for $\beta$ is
shown in the figure and one again sees explicit evidence for the
nonperturbative existence of a Landau pole. It is educational to compare
the Landau pole energy obtained from the exact result to the one loop estimate.
For example, with $\alpha(\mu^2 )=1.365$ we get
$\Lambda_{\rm Landau}^2 =2\mu^2$ while at one loop we get
$\Lambda_{\rm Landau}^2 =1.565\mu^2$. The value of $\alpha(\mu^2 )$ we
chose is very high because cutoff effects ought to be substantial when
$\mu$ is so close to $\Lambda_{\rm Landau}$. A top quark corresponding
to such a strong coupling would have a mass of $m_t =4\pi\sqrt{{\alpha}
\over{12}}~ 246~[GeV]\approx 1043~GeV$. We see that ``triviality bounds''
on the top will likely come out close to the perturbative unitarity bounds of
Chanowitz et. al. [5] a property of our
approximation that we view as an improvement on the work of
Einhorn and Goldberg [6] who obtained $5600 ~GeV$.
Our estimate is close to that obtained from simple variants of large
$N_w$($N_c$ fixed) or large $N_c$($N_w$ fixed) expansions [7]; this
agreement holds essentially because a
truncation of the set of planar diagrams to one loop turns out
to be relatively
acceptable numerically. From our non--perturbative result
in the figure one sees that the one and two loop errors in $\beta$
are about comparable for $\alpha\approx 0.4$ (roughly
of order 20 percent) and that using the one loop results
all the way out to infinity is not very harmful, but the two loop
result would be completely misleading if extrapolated beyond its
region of validity, to $\alpha\approx 1$. A top mass of $200~GeV$ would
correspond to $\alpha\approx 0.05$, well within the perturbative domain.

We hope to work out a more comprehensive analysis of the $N=\infty$
limit in the future, starting with a calculation of a $\beta$--function
associated with $\lambda_0$.
\vskip .5cm
\noindent {\bf Acknowledgement.}
This research was supported in part by the DOE under grant
\# DE-FG05-90ER40559.
\vskip 1cm
\noindent{\bf Figure Caption: }
The nonperturbatively determined $\beta$--function for the Yukawa coupling
at $N_c ,N_w \rightarrow\infty$ with ${{N_c}\over{N_w}}={3\over 2}$.
At tree level the top quark mass is given
by $m_t =4\pi\sqrt{{\alpha} \over{12}}~ 246~[GeV]$;
$\beta (\alpha ) = {{d\alpha} \over {dt}}$ where $t$ is
the logarithm of the energy scale.
\vskip 1cm

\leftline {\bf References}
\vskip .5cm

\item {[1]} H. Neuberger, Talk delivered  at the Topical Workshop
``Non perturbative aspects of chiral gauge theories'',
Accademia Nazionale dei Lincei, Roma, 9-11 March, 1992; to appear in the
proceedings. A more extensive list of references can be found here.

\item {[2]} M. Lindner, M. Sher and W. Zaglauter, {\bf Phys. Lett. B228}
(1989) 139.

\item {[3]} J. Bagger, S. Naculich, {\bf Phys. Rev. Lett. 67} (1991) 2252.

\item {[4]} R. Dashen and H. Neuberger, {\bf Phys. Rev. Lett. 50} (1983) 1897.

\item {[5]} M. S. Chanowitz, M. A. Furman and I. Hinchliffe, {\bf Nucl.
Phys. B153} (1979) 403.

\item {[6]} M. B. Einhorn and G. Goldberg, {\bf Phys. Rev. Lett. 57} (1986)
2115.

\item {[7]} K. Aoki, {\bf Phys. Rev. D44} (1991) 1547; S. Peris, {\bf
Phys. Lett. B251} (1990) 603.
\vfill
\eject
\end